\documentclass[conference]{IEEEtran}
\IEEEoverridecommandlockouts
\usepackage{cite}
\usepackage{amsmath,amssymb,amsfonts}
\usepackage{algorithm}
\usepackage{algpseudocode}
\usepackage{graphicx}
\usepackage{textcomp}
\usepackage{multirow} 
\usepackage{xcolor}
\def\BibTeX{{\rm B\kern-.05em{\sc i\kern-.025em b}\kern-.08em
    T\kern-.1667em\lower.7ex\hbox{E}\kern-.125emX}}
\begin{document}

\title{SSAAM: Sentiment Signal-based Asset Allocation Method with Causality Information \\
}

\author{\IEEEauthorblockN{Rei Taguchi}
\IEEEauthorblockA{\textit{School of Engineering} \\
\textit{The University of Tokyo}\\
Tokyo, Japan \\
s5abadiee@g.ecc.u-tokyo.ac.jp
}
\and
\IEEEauthorblockN{Hiroki Sakaji}
\IEEEauthorblockA{\textit{School of Engineering} \\
\textit{The University of Tokyo}\\
Tokyo, Japan \\
sakaji@sys.t.u-tokyo.ac.jp
}
\and
\IEEEauthorblockN{Kiyoshi Izumi}
\IEEEauthorblockA{\textit{School of Engineering} \\
\textit{The University of Tokyo}\\
Tokyo, Japan \\
izumi@sys.t.u-tokyo.ac.jp
}
}

\maketitle

\begin{abstract}
This study demonstrates whether financial text is useful for tactical asset allocation using stocks by using natural language processing to create polarity indexes in financial news.  In this study, we performed clustering of the created polarity indexes using the change-point detection algorithm. In addition, we constructed a stock portfolio and rebalanced it at each change point utilizing an optimization algorithm. Consequently, the asset allocation method proposed in this study outperforms the comparative approach. This result suggests that the polarity index helps construct  the equity asset allocation method.
\end{abstract}

\begin{IEEEkeywords}
Financial news, MLM scoring, causal inference, change-point detection, portfolio optimization 
\end{IEEEkeywords}

\section{Introduction}
\label{Introduction}

This study proposes that financial text can be useful for tactical asset allocation methods using equities. This study focuses on the point at which stock and portfolio prices change rapidly due to external factors, that is, the point of regime change. Regimes in finance theory refer to invisible market states, such as expansion, recession, bulls, and bears. In this study, we specifically drew on the two studies presented below. Wood et al.\cite{Wood_2021} used a change-point detection module to capture regime changes and created a simple and expressive model. Ito et al.\cite{itomasatake2021} developed a method for switching investment strategies in response to market conditions. In this study, we go one step further and focus on how to measure future regime changes. If the information on future regime changes (i.e., future changes in the market environment) is known, active management with a higher degree of freedom becomes possible. However, there are certain limitations in calculating future regimes using only traditional financial time-series data. Therefore, this study constructs an investment strategy based on a combination of alternative data that has been attracting attention in recent years and financial time-series data.

In this study, we hypothesized the following:

\begin{itemize}
\item Portfolio performance can be improved by switching between risk-minimizing and return-maximizing optimization strategies according to the change points created by the polarity index.
\end{itemize}

The contributions of this study are as follows:

\begin{itemize}
\item We demonstrate that the estimation of regime change points using financial text is active the active management and propose a highly expressive asset allocation framework.
\end{itemize}

The framework of this study consists of the following four steps.

\begin{itemize}
\item \textbf{Step 1 (Creating polarity index): }Score financial news titles using MLM scoring. In addition, quartiles are calculated from the same data, and a three-value classification of positive, negative, and neutral is performed according to the quartile range. The calculated values are aggregated daily.
\end{itemize}

\begin{itemize}
\item  \textbf{Step 2 (Demonstration of leading effects): }We use statistical causal inference to demonstrate whether financial news has leading effects on a stock portfolio. Use the polarity index created in Step 1. We will also create a portfolio of 10 stocks combined. The algorithm used is VAR-LiNGAM.
\end{itemize}

\begin{itemize}
\item  \textbf{Step 3 (Change point detection): }Verify that the polarity index has leading effects in Step 2. Calculate the regime change point of the polarity index using the change point detection algorithm. The algorithm used is the Binary Segmentation Search Method.
\end{itemize}

\begin{itemize}
\item  \textbf{Step 4 (Portfolio optimization): }Portfolio optimization is performed based on the change points created in Step 3. The algorithm used is EVaR optimization.
\end{itemize}

\section{Method}
\label{Method}

\subsection{Creating polarity index}
\label{Creating polarity index}
This study used pseudo-log-likelihood scores (PLLs) to create polarity indices. PLLs are scores based on probabilistic language models proposed by Salazar et al.\cite{salazar-etal-2020-masked}. Because masked language models (MLMs) are pre-trained by predicting words in both directions, they cannot be handled by conventional probabilistic language models. However, PLLs can determine the naturalness of sentences at a high level because they are represented by the sum of the log-likelihoods of the conditional probabilities when each word is masked and predicted. Token $\psi_t$ is replaced by [MASK], and the past and present tokens $\textbf{$\Psi$}_{\backslash t}= [\psi_1,\psi_2,...,\psi_{t}]$ are predicted. $t$ represents time. $\Theta$ is the model parameter. $P_{MLM} (\cdot)$ denotes the probability of each sentence token. The MLM selects BERT (Devlin et al.\cite{devlin2019bert}).

\begin{align}
\label{PLL}
\textbf{PLL($\Psi$)} := \sum^{|\textbf{$\Psi$}|}_{t=1} \log_{2} P_{MLM} (\psi_t | \textbf{$\Psi$}_{\backslash t}; \Theta)
\end{align}

After pre-processing, score the financial news text with PLLs one sentence at a time. Quartile ranges\footnote{Arranging the data in decreasing order, the data in the 1/4 are called the 1st quartile, the data in the 2/4 are called the 2nd quartile, and the data in the 3/4 are called the 3rd quartile.  (3rd quartile - 1st quartile) is called the quartile range.} were calculated for data that scored one sentence at a time. The figure below illustrates the polarity classification method.

\begin{table}[htbp]
\centering
  \caption{Polarity Classification Method}
  \label{tb:Polarity Judgment Method}
  \scalebox{1}[1]{ 
  \begin{tabular}{cc}  \hline
    \multicolumn{1}{c}{Classification Method}  & \multicolumn{1}{c}{Sentiment Score} \\ \hline \hline
    3rd quartile $<$ PLLs & 1 (positive)  \\ 
    1st quartile $\leq$ PLLs $\leq$ 3rd quartile & 0 (neutral)  \\ 
    1st quartile $>$ PLLs & -1 (negative)  \\   \hline 
  \end{tabular}
  }
\end{table}

Aggregate the scores chronologically according to the title column of financial news.

\subsection{Demonstration of leading effects}
\label{Demonstration of leading effects}
In this study, we used VAR-LiNGAM to demonstrate the precedence. VAR-LiNGAM is a statistical causal inference model proposed by Hyv\"arinen et al.\cite{hyvarinen2010estimation}. The causal graph inferred by VAR-LiNGAM is as follows:

\begin{align}
\label{VAR-LiNGAM}
\textbf{x}(t) = \sum^{T}_{\tau=1} \textbf{B}_{\tau} \textbf{x}(t-\tau)+\textbf{e}(t)
\end{align}

where $\textbf{x}(t)$ is the vector of the variables at time $t$ and $\tau$ is the time delay. $T$ represents the maturity date. In addition, $\textbf{B}_{\tau}$ is a coefficient matrix that represents the causal relationship between the variables $\textbf{x}(t-\tau)$. $\textbf{e}(t)$ denotes the disturbance term. VAR-LiNGAM was implemented using the following procedure: First, a VAR (Vector Auto-Regressive) model is applied to the causal relationships among variables from the lag time to the current time. Second, for the causal relationships among variables at the current time, LiNGAM inference is performed using the residuals of the VAR model. This study confirms whether financial news is preferred to  stock portfolios.

\subsection{Change point detection}
\label{Change point detection}
Binary segmentation search (Bai\cite{bai1997estimating};  Fryzlewicz\cite{fryzlewicz2014wild}) is a greedy sequential algorithm. The notation of the algorithm follows Truong et al.\cite{truong2020selective}. This operation is greedy in the sense that it seeks the change point with the lowest sum of costs. Next, the signal was divided into two at the position of the first change point, and the same operation was repeated for the obtained partial signal until the stop reference was reached. The binary segmentation search is expressed in Algorithm \ref{alg:Binary Segmentation Search}. We define a signal $y=\{y_s\}^S_{s=1}$ that follows a multivariate non-stationary stochastic process. This process involves $S$ samples. $L$ refers to the list of change points. Let $s$ denote the value of a change point. $G$ refers to an ordered list of change points to be computed. If signal $y$ is given, the $(b-a)$-sample long sub-signal $\{y_s\}^b_{s=a+1}, (1 \leq a < b \leq S)$ is simply denoted $y_{a, b}$. Hats represent the calculated values. Other notations are noted in the algorithm’s comment.

\begin{algorithm}
\caption{Binary Segmentation Search}
\label{alg:Binary Segmentation Search}
\begin{algorithmic}
\renewcommand{\algorithmicrequire}{\textbf{Input:}}
\Require signal $y=\{y_s\}^S_{s=1}$, cost function $c(\cdot)$, stopping criterion.\\
Initialize $L \leftarrow \{ \}.$ \Comment{Estimated breakpoints}.
\renewcommand{\algorithmicrequire}{\textbf{Repeat}}
\Require
\State $k \leftarrow |L|.$ \Comment{Number of breakpoints}.
\State $s_0 \leftarrow 0$ and $s_{k+1} \leftarrow S$ \Comment{Dummy variables}
\If{$k > 0$}
\State Denote by $s_i (i=1,...,k)$ the elements (in ascending order) of $L$, ie $L=\{s_1,...,s_k\}.$   
\EndIf\\
Initialize $G$ a $(k + 1)$-long array. \Comment{List of gains}
\For {$i = 0,...,k$}
\State $G[i] \leftarrow c(y_{s_i, s_{i+1}})-\mathop{\min}_{s_i<s<s_{i+1}} [c(y_{s_i, s})+c(y_{s, s_{i+1}})].$
\EndFor\\
$\hat{i} \leftarrow \mathop{\arg \max}_i G[i]$\\
$\hat{s} \leftarrow \mathop{\arg \min}_{s_i<s<s_{i+1}} [c(y_{s_{\hat{i}}, t})+c(y_{s, s_{{\hat{i}}+1}})].$\\ \Comment{Estimated change-points }\\
$L \leftarrow L \cup \{\hat{s}\}$ 
\renewcommand{\algorithmicrequire}{\textbf{Until}}
\Require stopping criterion is met.
\renewcommand{\algorithmicensure}{\textbf{Output:}}
\Ensure set $L$ of estimated breakpoint indexes.
\end{algorithmic}
\end{algorithm}

.

\subsection{Portfolio optimization}
\label{Portfolio optimization}

The entropy value at risk (EVaR) is a coherent risk measure that is the upper bound between the value at risk (VaR) and conditional value at risk (CVaR) derived from Chernoff's inequality (Ahmadi-Javid\cite{6033932}; Ahmadi-Javid\cite{ahmadi2012entropic}). EVaR has the advantage of being computationally tractable compared to other risk measures, such as CVaR, when incorporated into stochastic optimization problems (Ahmadi-Javid\cite{ahmadi2012entropic}). EVaR is defined as follows.

\begin{align}
\label{EVaR}
\textbf{EVaR}_{\alpha}(X):=\min_{z>0}\left\{z \ln \left(\frac{1}{\alpha} M_{X} \left(\frac{1}{z}\right) \right)\right\}
\end{align}

$X$ is a random variable. $M_{X}$ is the moment-generating function. $\alpha$ denotes the significance level. $z$ are variables. A general convex programming framework for the EVaR is proposed by Cajas\cite{cajas2021entropic}. In this study, we switch between the following two optimization strategies depending on the regime classified in Section \ref{Change point detection}.

\begin{itemize}
\item \textbf{Minimize risk optimization: }A convex optimization problem with constraints imposed to minimize EVaR given a level of expected $\mu$ ($\widehat{\mu})$.
\end{itemize}

\begin{align}
\label{MinRisk}
      \begin{aligned}
          & \text{minimize}
              & q+z \log_{e} \left(\frac{1}{T \alpha}\right)\\
          & \text{subject to}
              & \mu w \ge \widehat{\mu} \\           
              & & \sum^N_{i=1} w_i = 1\\
              & & z \ge \sum^T_{j=1} u_j\\
              & & (-r_jw^\top -q, z, u_j) \in{K_{exp}} 
              & &(\forall j=1,...,T)\\
              & & w_i = 0
              & &(\forall i=1,...,N)\\
      \end{aligned}
\end{align}

\begin{itemize}
\item \textbf{Maximize return optimization: }A convex optimization problem imposed to maximize expected return given a level of expected $EVaR$ ($\widehat{EVaR}$).
\end{itemize}

\begin{align}
\label{MaxRet}
      \begin{aligned}
          & \text{maximize}
              & \mu w^\top \\  
          & \text{subject to}
              & q+z \log_{e} \left(\frac{1}{T \alpha}\right) \ge \widehat{EVaR}\\ 
              & & \sum^N_{i=1} w_i = 1\\
              & & z \ge \sum^T_{j=1} u_j\\
              & & (-r_jw^\top -q, z, u_j) \in{K_{exp}} 
              & &(\forall j=1,...,T)\\
              & & w_i = 0
              & &(\forall i=1,...,N)\\
      \end{aligned}
\end{align}
   
where $q$, $z$ and $u$ are the variables, $K_{exp}$ is the exponential cone, and $T$ is the number of observations. $w$ is defined as a vector of weights for $N$ assets, $r$ is a matrix of returns, and $\mu$ is the mean vector of assets.

\section{Experiments \& Results}
\label{Experiments & Results}

\subsection{Dataset description}
\label{Dataset description}
This study calculates the signal for portfolio rebalancing and tactical asset allocation to actively go for an alpha based on the assumption that financial news  precedes the equity portfolio. Two types of data were used.

\begin{itemize}
\item \textbf{Stock Data: }We used the daily stock data provided by Yahoo!Finance\footnote{https://finance.yahoo.com/}. The stocks used are the components of the NYSE FANG+ Index: Facebook, Apple, Amazon, Netflix, Google, Microsoft, Alibaba, Baidu, NVIDIA, and Tesla were selected. For this data, adjusted closing prices are used. The time period for this data is January 2015 through December 2019.
\end{itemize}

\begin{itemize}
\item \textbf{Financial News Data: }We used the daily historical financial news archive provided by Kaggle\footnote{https://www.kaggle.com/}, a data analysis platform. This data represents the historical news archive of U.S. stocks listed on the NYSE/NASDAQ for the past 12 years. This data was confirmed to contain information on ten stock data issues. This data consists of 9 columns and 221,513 rows. The title and release date columns were used in this study. The time period for this data is January 2015 through December 2019.
\end{itemize}

\subsection{Preparation for backtesting}
\label{Preparation for backtesting}
The polarity index is presented in section \ref{Creating polarity index}. The financial news data were pre-processed once before creating the polarity index. Both financial news and stock data are in daily units; however, to match the period, if there are blanks in either , lines containing blanks are dropped. Once the polarity index is created in Section \ref{Creating polarity index}, the next step is to create a stock portfolio by adding the adjusted closing prices of 10 stocks. The investment ratio for the portfolio is set uniformly for all stocks. Next, we use VAR-LiNGAM in Section \ref{Demonstration of leading effects} to perform causal inference. The causal inference results are as follows: Python library ruptures (Truong et al.\cite{truong2020selective}) was used.

\begin{table}[htbp]
\centering
  \caption{Causal Inference in VAR-LiNGAM}
  \label{tb:Causal Inference in VAR-LiNGAM}
  \scalebox{1}[1]{ 
  \begin{tabular}{cc}  \hline
    \multicolumn{1}{c}{Direction}  & \multicolumn{1}{c}{Causal Graph Value} \\ \hline \hline
    Index(t-1) $\dashrightarrow$ Index(t) & 0.39  \\ 
    Index(t-1) $\dashrightarrow$ Portfolio(t) & \textbf{0.11}  \\ 
    Portfolio(t-1) $\dashrightarrow$ Portfolio(t) & 1.00  \\   
    \hline 
  \end{tabular}
  }
\end{table}

The values in Table \ref{tb:Causal Inference in VAR-LiNGAM} refer to the elements of the adjacency matrix. The lower limit was set to 0.05. The results in the table show that the polarity index has a leading edge in the equity portfolio. The Python library LiNGAM (Hyv\"arinen et al.\cite{hyvarinen2010estimation}) was used.

\subsection{Backtesting scenarios}
\label{Backtesting scenarios}

In this study, the following rebalancing timings were merged and backtested. Python library vector (Polakow\cite{vectorbt}) and Riskfolio-Lib (Cajas\cite{riskfolio}) was used for backtesting. In addition to EVaR optimization, CVaR optimization and the mean-variance model were used as optimization algorithms and comparative methods, respectively. In this study, the number of regimes was set to 5 and 10. The rebalancing times were 30, 90, and 180 days. The backtesting methodology was as follows. In this study, CPD-EVaR++ was positioned as the proposed strategy, and CPD-EVaR+ was the runner-up strategy.

\begin{itemize}
\item \textbf{CPD-EVaR++ (proposed): }Changepoint rebalancing using risk minimization and return maximization EVaR optimization + regular intervals rebalancing strategy
\item \textbf{CPD-EVaR+: }Changepoint rebalancing using risk minimization and no-restrictions EVaR optimization + regular intervals rebalancing strategy
\item \textbf{EVaR: }EVaR optimization regular intervals rebalancing strategy
\item \textbf{CVaR: }CVaR optimization regular intervals rebalancing strategy
\item \textbf{MV: }Mean-Variance optimization regular intervals rebalancing strategy
\end{itemize}

The binary determination of whether the polarity index within each regime shows an upward or downward trend is made by examining the divided regimes. MinRiskOpt (Section $\ref{Portfolio optimization}-(\ref{MinRisk})$) is assigned to an upward trend, and MaxReturnOpt (Section $\ref{Portfolio optimization}-(\ref{MaxRet})$) is assigned to a downward trend.

\subsection{Evaluation by backtesting}
\label{Evaluation by backtesting}

The following metrics were employed to assess the portfolio performance.

\begin{itemize}
\item \textbf{Total Return (TR): }TR refers to the total return earned from investing in an investment product within a given period. TR formula is as follows: TR = Valuation Amount + Cumulative Distribution Amount Received + Cumulative Amount Sold - Cumulative Amount Bought. This study does not incorporate tax amounts and trading commissions. 
\item \textbf{Maximum Drawdown (MDD): }MDD refers to the rate of decline from the maximum asset. MDD formula is as follows: MDD = (Trough Value - Peak Value) / Peak Value.
\end{itemize}

\begin{table}[htbp]
    \centering
    \caption{Backtesting (SSAAM)}
    \label{tb:Backtesting (SSAAM)}
  \begin{tabular}{ccccc}  \hline
    \multicolumn{1}{c}{Rebalance} & \multicolumn{1}{c}{Regime} & \multicolumn{1}{c}{Algorithm} & \multicolumn{1}{c}{TR \lbrack \%\rbrack} & \multicolumn{1}{c}{MDD \lbrack \%\rbrack}\\ \hline \hline
    \multirow{4}{*}{30-days} & \multirow{2}{*}{5} & CPD-EVaR++ & \textbf{810.9915}  & 26.8629\\
    & & CPD-EVaR+ & 594.7410  & 26.8629\\
    \cline{2-5}
    & \multirow{2}{*}{10} & CPD-EVaR++ & 485.5201 & 45.0235\\
    & & CPD-EVaR+ & 392.1392  & 42.4803\\
    \hline
    \multirow{4}{*}{90-days} & \multirow{2}{*}{5} & CPD-EVaR++ & 535.7349  & 27.6386 \\
    & & CPD-EVaR+ & 410.8530  & 27.6386 \\
    \cline{2-5}
    & \multirow{2}{*}{10} & CPD-EVaR++ & 417.8354  & 27.7646 \\
    & & CPD-EVaR+ & 373.5849  & 27.7646 \\
    \hline
    \multirow{4}{*}{180-days} & \multirow{2}{*}{5} & CPD-EVaR++ & 152.0988  & 27.3924 \\
    & & CPD-EVaR+ & 131.2210  & 27.3924 \\
    \cline{2-5}
    & \multirow{2}{*}{10} & CPD-EVaR++ & 169.2992 & \textbf{25.3050} \\
    & & CPD-EVaR+ & 232.4513  & \textbf{25.3050} \\
    \hline 
      \end{tabular}
\end{table}

\begin{table}[htbp]
    \centering
    \caption{Backtesting (comparison)}
    \label{tb:Backtesting (comparison)}
    \begin{tabular}{cccc}  \hline
    \multicolumn{1}{c}{Rebalance} &  \multicolumn{1}{c}{Algorithm} &  
    \multicolumn{1}{c}{TR \lbrack \%\rbrack} & \multicolumn{1}{c}{MDD \lbrack \%\rbrack}\\ 
    \hline \hline
    \multirow{3}{*}{30-days}  & EVaR & \textbf{587.9630}  & 46.6651\\
    & CVaR & 558.7446  & 44.4532\\
     & MV & 527.2827  & 42.9851\\
    \hline 
    \multirow{3}{*}{90-days} & EVaR & 500.1421  & 44.9860\\
    & CVaR & 496.7423 & 44.0592\\
    & MV & 459.1195  & 42.7358\\
    \hline 
    \multirow{3}{*}{180-days} & EVaR & 353.2412  & 44.7714 \\
    & CVaR & 382.9451  & 44.2525\\
    & MV & 360.4298  & \textbf{42.8165}\\
    \hline
      \end{tabular}
\end{table}

\section{Discussion \& Conclusion}

Table \ref{tb:Backtesting (SSAAM)} shows that the higher the number of regular rebalances, the higher the total return. In addition, the maximum drawdowns hovered between 25\% and 45\%, which is considered acceptable to the average system trader. In this study, the experiment was conducted separately when the regime was five and when the regime was ten. The total return was higher when the regime was five, whereas the maximum drawdown was almost the same for both regimes. Moreover, as hypothesized, CPD-EVaR++, a combination of risk minimization and return maximization operations, performed better than the others. Therefore, using this method, the best practice in managing equity portfolios is to use CPD-EVaR++ and to rebalance irregularly in regime 5, in addition to regular rebalancing every 30 days.

Backtesting of Table \ref{tb:Backtesting (comparison)} using the same parameters as in Table \ref{tb:Backtesting (SSAAM)}. The results show that for the algorithm, EVaR optimization performed better than the others, similar to the results of Cajas\cite{cajas2021entropic}. This may be because the computational efficiency of EVaR in stochastic optimization problems is higher than that of other risk measures, such as CVaR.

This study demonstrates the utility of financial text in asset allocation with equity portfolios. In the future, we would like to develop a tactical asset allocation strategy that mixes stocks and other asset classes, such as bonds. In the future, we would also like to apply this research to monetary policy and other macroeconomic analyses.

\section*{Acknowledgment}
This work was supported by the JST-Mirai Program Grant Number JPMJMI20B1, Japan. The authors declare that the research was conducted without any commercial or financial relationships that could be construed as potential conflicts of interest.

\bibliographystyle{unsrt}
\bibliography{ieee}

\vspace{12pt}

\end{document}